\documentclass[aps,prd,twocolumn]{revtex4-1}
\usepackage{}
\usepackage{mathrsfs}
\usepackage{exscale}                  
\usepackage[intlimits]{amsmath}       
\usepackage{amsfonts}
\usepackage{amssymb,amscd}
\usepackage[dvips]{epsfig}
\usepackage{array}
\usepackage{wrapfig}
\usepackage{color}
\newcommand{\beqa}{\begin{eqnarray}}
\newcommand{\eeqa}{\end{eqnarray}}
\newcommand{\beq}{\begin{equation}}
\newcommand{\eeq}{\end{equation}}

\begin{document}
\title{A thermodynamically consistent quasi-particle model without temperature-dependent infinity of the vacuum zero point energy}
\author{Jing Cao$^{1}$, Yu Jiang$^{2}$, Wei-min Sun$^{1,3}$, Hong-shi Zong$^{1,3}$}
\address{$^{1}$ Department of Physics, Nanjing University, Nanjing 210093, China}
\address{$^{2}$ College of Mathematics, Physics and Information Engineering, Zhejiang Normal University, Jinhua 321004, China}
\address{$^{3}$ Joint Center for Particle, Nuclear Physics and Cosmology, Nanjing 210093, China}
\date{\today}

\begin{abstract}

In this paper, an improved quasi-particle model is presented. Unlike the previous approach of establishing quasi-particle model, we introduce a classical background field (it is allowed to depend on the temperature) to deal with the infinity of thermal vacuum energy which exists in previous quasi-particle models. After taking into account the effect of this classical background field, the partition function of quasi-particle system can be made well-defined. Based on this and following the standard ensemble theory, we construct a thermodynamically consistent quasi-particle model without the need of any reformulation of statistical mechanics or thermodynamical consistency relation. As an application of our model, we employ it to the case of (2+1) flavor QGP at zero chemical potential and finite temperature and obtain a good fit to the recent lattice simulation results of S. Borsanyi $et$ $al$. A comparison of the result of our model with early calculations using other models is also presented. It is shown that our method is general and can be generalized to the case where the effective mass depends not only on the temperature but also on the chemical potential.

\bigskip

E-mail: zonghs@chenwang.nju.edu.cn. ~~~~PACS numbers: 12.38.Mh, 51.30.+i, 52.25.Kn.

\end{abstract}

\maketitle

\section{Introduction}

It is generally accepted that the theory describing strongly interacting matter is Quantum Chromodynamics (QCD). In QCD with two massless quarks, 
a spontaneously broken chiral symmetry is restored at finite temperature and zero chemical potential. It can be argued \cite{Pisarski, Rajagopal} that this phase transition is likely second order. If this transition is indeed second order, QCD with two quarks of nonzero mass has only a smooth crossover as a function of $T$. Although not yet firmly established, this picture is consistent with present lattice simulations and many models. Meanwhile, at finite temperature and zero chemical potential, the transition from the confined hadronic matter to the deconfined quark-gluon plasma (QGP) is understood now as a 
crossover for QCD in the real world \cite{Aoki,Bazavov,Aoki1}.
Because QGP describes the relevant features of nature under extreme conditions, for example, in neutron stars (at large density \cite{28}), or in the early universe (at essentially vanishing net baryon density \cite{26}), the study of thermodynamical properties of QGP has attracted considerable attention over the years from the theory community. In principle, lattice QCD provides a straightforward way to calculate properties of QGP, and in particular, its equation of state (EOS). The techniques of these calculations have been developed greatly in the last years, even for the case of finite chemical potentials \cite{1,2,3,4,5}. Among these analytical approaches, the weak-coupling expansion of thermodynamical quantities shows a extremely poor convergence for any temperature of practical interest \cite{27}. To overcome this poor convergence, researchers have invented many methods with a rigorous link to QCD, notably the resummed HLT scheme \cite{16,17}, $\Phi$-functional approach \cite{18,19}. On the other hand, many different phenomenological models were adopted to describe the non-perturbative behavior of QGP seen in lattice simulations of QCD. Among those models, the quasi-particle quark-gluon plasma model with a few fitting parameters, has been widely used to reproduce the properties of the QCD plasma \cite{9,8,10,14,20,11,15,12,22,13,31}. In this model, at finite temperature, instead of real quarks and gluons with QCD interactions, one can consider the system to be made up of non-interacting quasi-particles, i.e., quasi-quarks and quasi-gluons, with temperature-dependent effective mass. 
The quasi-particle model was first proposed by Goloviznin and Satz \cite{9}, and later by Peshier $et$ $al$ \cite{8}. After some time, Gorenstein and Yang found that this model was thermodynamically inconsistent, and remedied this flaw by reformulating statistical mechanics \cite{10}. After that, Bannur also proposed a new quasi-particle model using standard statistical mechanics and avoided the thermodynamical inconsistency from the energy density rather than the pressure \cite{13}. Furthermore, in Ref. \cite{29} Gardim and Steffens showed that the two models proposed by Peshier and Bannur were two extreme limits of a general formulation. 

Over the past few years, more considerable progress has been made in the quasi-particle model. However, just as will be shown below, in all previous works on quasi-particle model  the problem of the temperature-dependent infinity of the vacuum zero point energy and its influences have not been seriously considered. In a consistent quasi-particle model, how to eliminate this temperature-dependent infinity of the vacuum zero point energy is very important. In the present work we try to answer this question.

The outline of this paper is as follows. In section II, we show the problem existed in previous quasi-particle model. In sections III and IV, adopting the series expansion method inspired by Walecka to deal with the temperature-dependent infinity of the vacuum zero point energy, we construct a new thermodynamically consistent framework of quasi-particle model for QGP without the need of any reformulation of statistical mechanics or thermodynamical consistency relation.  As an application of our model, we employ it to the case of (2+1) flavor QGP at zero chemical potential and finite temperature to fit the recent lattice simulation results and compare our results with those of earlier models.  In section V, we conclude our work with a summary of the results.

\section{Problems in previous models}

To construct our quasi-particle model, we first illustrate the problem hidden in early quasi-particle models. For simplicity, we take the scalar field as an example, similar problems also exist in the case of fields with spin. As is well known, one always uses the scalar field to characterize a system composed of Bose-type quasi-particles. Now, let us begin with the Lagrangian of quasi-particle with a temperature-dependent mass $m=m(T)$
\begin{equation}
\mathscr{L}=\sum\limits_{i=1}^{N}\frac{1}{2}\partial_{\mu}\phi_{i}\partial^{\mu}\phi_{i}-\frac{1}{2}m^{2}(T)\phi_{i}^{2},
\end{equation}
where the effective mass $m(T)$ describes the interaction between real particles and $N$ is the number of species of quasi-particles. Here, following Ivanov et.al~\cite{21}, we use this effective Lagrangian to describe an interacting system composed of gluons, provided that we assume that there are only massive and transverse quasi-gluons.

Inserting the Fourier expansion of the field $\phi_{i}(x)$
\begin{displaymath}
\phi_{i}(x)=\sqrt{\frac{\beta}{V}}\sum\limits_{n=-\infty}^{\infty}\sum_{\vec{p}}
\exp{[i(\vec{p}\cdot\vec{x}+\omega_{n}\cdot\tau)]\phi_{i_{ n}}(\vec{p})}
\end{displaymath}
into this Lagrangian and according to the definition of partition function in the path integral formalism \cite{23,24}, one obtains
\begin{displaymath}
\begin{split}
Z&=\mathrm{Tr}e^{-\beta H}\\&=N'\prod\limits_{i=1}^{N}\int\limits_{periodic}\mathscr{D} \phi_{i}\exp\bigg[\int\limits_{0}^{\beta}d\tau\int d^3x\mathscr{L}\bigg].
\end{split}
\end{displaymath}
Here the above formula expresses the partition function $Z$ as a functional integral over $\phi_{i}$ of the exponential of the action in imaginary time and $N^{\prime}$ is an irrelevant overall normalization constant. Then, using the standard path integral procedure, we have the following partition function:
\begin{equation}
\ln Z=VN\int\frac{d^3p}{(2\pi)^3}\bigg[-\frac{1}{2}\beta \omega^{\ast}-\ln(1-e^{-\beta \omega^{\ast}})\bigg],
\end{equation}
where $\omega^{\ast}=\sqrt{p^{2}+m^{2}(T)}$ is the dispersion relation for the quasi-particle. From this expression we can see that the second term of the right-hand-side of Eq. (2) has the form of the ideal Bose gas formula for a quasi-particle except for the temperature-dependent dispersion relation coming from the effective mass $m(T)$; the first term of the right-hand-side of Eq. (2) is the zero point energy of the ``vacuum'' and is divergent. The difference between our case and that of standard statistical mechanics is that the infinity depends on the temperature $T$. In other words, the ``vacuum'' in our case is a thermal vacuum. In standard statistical mechanics we have $m=const$,
and that divergence is independent of the temperature $T$, so we can throw this divergent part away in this case (this is because the vacuum zero-point energy and pressure cannot be measured experimentally and therefore the zero-point energy should be subtracted).  
This operation will have no effect on computation of thermodynamical quantities. For example, by taking the derivative of the partition function, we can obtain the equation of state and the energy density:
\begin{displaymath}
\mathcal {P}=\frac{1}{\beta}\frac{\partial\ln Z}{\partial\ V}=-T\frac{d}{2\pi^{2}}\int\limits_{0}^{\infty}{p^{2}\ln(1-e^{-\frac{\omega}{T}})d p},
\end{displaymath}
\begin{displaymath}
\mathcal {E}=-\frac{1}{V}\frac{\partial\ln Z}{\partial\ \beta}=\frac{d}{2\pi^{2}}\int\limits_{0}^{\infty}{p^{2}\frac{\omega}{e^{\frac{\omega}{T}}-1}d p},
\end{displaymath}
where $\omega=\sqrt{p^{2}+m^{2}}$ and $d=N$ is the degeneracy factor ($d=6$ for SU(2); and $d=16$ for SU(3) gluons). The above formula is formally model independent because it comes completely from ensemble theory. We can see that the divergent part vanishes automatically after taking the derivative, because it is only an infinite constant. Now let us turn back to our current effective Lagrangian. When the function $\omega^{\ast}(p)$ for the particle (``quasi-particle'' in our case) excitation energy becomes temperature-dependent, the operation of taking the derivative with respect to $\beta$ will no longer be valid. In particular, in the definition of energy density, we find there is an extra term coming from the divergent part. If one wants to obtain  physically meaningful results, he or she must treat this infinity carefully rather than throwing it away naively. A similar problem also emerges in the case of Fermi fields. For spin $1/2$ field at zero chemical potential, from the Lagrangian of the quasi-fermion
\begin{equation}
\mathscr{L}=\bar{\psi}(i\gamma^{\mu}\partial_{\mu}-m(T))\psi,
\end{equation}
one can do the derivation as before, and finally obtain:
\begin{equation}
\ln Z=2V\int\frac{d^3p}{(2\pi)^3}\bigg[\beta\omega^{\ast}+2\ln(1+e^{-\beta\omega^{\ast}})\bigg],
\end{equation}
where $\omega^{\ast}=\sqrt{p^{2}+m^{2}(T)}$ is the dispersion relation for a Fermi-type quasi-particle. Similar to the case of Bose-type quasi-particle discussed above, there is also a temperature-dependent divergent term, which represents a temperature-dependent thermal vacuum energy. Indeed, just as was shown above, the two partition functions in both cases are ill-defined. If the infinity in the temperature-dependent thermal vacuum energy is not tackled properly, we will not be able to get physically meaningful results in the quasi-particle model. The aim of the rest of this paper is to treat this temperature-dependent ``vacuum'' energy carefully and establish a thermodynamically consistent quasi-particle model.

\section{Elimination of thermal vacuum divergence}

In this section, we do our calculation for Dirac field at finite temperature and zero chemical potential and then turn to the case of scalar field. To eliminate the temperature-dependent infinity of the thermal vacuum energy, we introduce a classical background field $B$ (it is allowed to depend on the temperature) into the Lagrangian of the quasi-particle system. Thus, for a system composed of Fermi-type quasi-particles, we have the following Lagrangian
\begin{equation}
\mathscr{L}=\bar{\psi}(i\gamma^{\mu}\partial_{\mu}-m(T))\psi+B.
\label{fermilag}
\end{equation}
Inspired by the approach of Walecka \cite{25}, we shall use a 
series expansion method to separate the divergence term, and then choose a appropriate classical background field $B$ that satisfies the following condition
\begin{equation}
\frac{B}{m_{0}^{4}}=-\mathrm{Tr}\int\frac{d^4k}{(2\pi)^4}i\sum\limits_{n=1}^{4}\frac{(\eta-1)^n}{n(\not\!k-1)^n},
\label{bagconstant}
\end{equation}
where $\eta=\frac{m(T)}{m_{0}}$, as the counter term to remove the divergence and make the shift in the ground-state energy of the total system finite.

After taking into account the effect of this classical background field, by means of dimensional regularization we will have eliminated the divergence of vacuum zero point energy and obtain the finite, physically meaningful result of the shift in the ground-state energy
\begin{eqnarray}
\Delta\varepsilon_{0}&&=E_{0}-B-E_{vac}\nonumber\\
&&=-\frac{2}{(4\pi)^2}\bigg[m^{4}(T)\ln\frac{m(T)}{m_{0}}+m_{0}^{3}(m_{0}-m(T))\nonumber\\
&&-\frac{7}{2}m_{0}^{2}(m_{0}-m(T))^{2}+\frac{13}{3}m_{0}(m_{0}-m(T))^{3}\nonumber\\
&&-\frac{25}{12}(m_{0}-m(T))^{4}\bigg].
\end{eqnarray}
This is what one expects in advance. Here we would like to stress that the series expansion method inspired by Walecka's approach for eliminating the temperature dependent infinity of vacuum zero point energy given in Eq. (6) is the simplest way to achieve this goal.

After completing the elimination of the infinity of thermal vacuum energy successfully in the Dirac field case, we turn our attention to the case of the scalar field. Similarly we start from the Lagrangian for Bose-type quasi-particle
\begin{equation}
\mathscr{L}=\frac{1}{2}\partial_{\mu}\phi\partial^{\mu}\phi-\frac{1}{2}m^{2}(T)\phi^{2}+B,
\label{g}
\end{equation}
where $m(T)$ is the effective mass of the quasi-particle. Using again the definition of measurable physical quantity and the series expansion, we can get the form of the classical background field
\begin{equation}
\frac{B}{m_{0}^{4}}=\int\frac{d^4k}{(2\pi)^4}i\frac{1}{2}\sum\limits_{n=1}^{2}\frac{(\eta^{2}-1)^n}{n(k^{2}-1)^n},
\end{equation}
where $\eta^{2}=\frac{m^{2}(T)}{m_{0}^{2}}$.
Owing to each boson's propagator having one more power of momentum in the denominator than that of the fermion, the classical background field $B$ in a Bose-type quasi-particle system has only two terms. As is done previously for the Dirac field case, to calculate this integral, we generalize it to $d$ dimensions and rotate to Euclidean space.  We perform the integration and take the limit. In the end we get a finite shift in the ground-state energy
\begin{eqnarray}
\Delta\varepsilon_{0}=\frac{1}{(4\pi)^2}\bigg[&&\frac{1}{2}m^{4}(T)\
\ln\frac{m(T)}{m_{0}}+\frac{1}{2}m_{0}^{2}m^2(T)\nonumber\\
&&-\frac{3}{8}m^4(T)-\frac{1}{8}m_{0}^{4}\bigg].
\end{eqnarray}
Here we remark that the last term on the right-hand-side of 
Eq. (10) is a constant, and can be thrown away with no effect. Thus, 
with the help of the background field, we obtain a well-defined partition function of the quasi-particle system.

The analysis of the divergence of vacuum zero point energy related to massless gauge bosons will be similar to the  previous one, but the result is different compared to the case where the rest mass of the quasi-particle is nonzero. We repeat the above analysis and arrive at the form of the classical background field for gauge bosons:
\begin{displaymath}
B=N\int\frac{d^4k}{(2\pi)^4}\frac{i}{2}\sum\limits_{n=1}^{2}(-1)^{n+1}\frac{m^{2n}(T)}{n(k^{2}-m^{2}(T))^{n}},
\end{displaymath}
where $N$ is the degeneracy factor ($N=6$ for SU(2); and $N=16$ for SU(3)). Here we still adopt dimensional regularization to calculate $E-E_{vac}$ and $B$. After some algebra we find:
\begin{equation}
\Delta\varepsilon_{0}=E_{0}-E_{vac}-B=\frac{N}{(4\pi)^2}\frac{1}{8}m^{4}(T).
\end{equation}
Therefore, in the case of gauge bosons, there is only one term proportional to the fourth power of thermal mass contributing to the finite shift in the ground-state energy.

\section{A self-consistent quasi-particle model}
We now use the quasi-bosons system as an example to illustrate the self-consistent statistical model for quasi-particles constructed above. It is well known that in equilibrium statistical field theory, once the partition function of the system is obtained, then all the thermodynamical variables can be determined. Therefore, the calculation of the system's partition function is very important.
This is the reason why in this paper we take the partition function as the starting point for studying the quasi-particle model.

For Bose-type quasi-particle, the partition function is
\begin{equation}
\ln Z=-V\beta\Delta\varepsilon_{0}-V\int\frac{d^3p}{(2\pi)^3}\ln(1-e^{-\beta \omega^{\ast}}),
\end{equation}
where $\Delta\varepsilon_{0}$ is the shift in the ground-state energy calculated in the last section for the scalar field. Then, according to the standard ensemble theory, we have
\begin{displaymath}
\mathcal {P}=\frac{1}{\beta}\frac{\partial\ln Z}{\partial V}=-\Delta\varepsilon_{0}-T\int\frac{d^3p}{(2\pi)^3}
\ln(1-e^{-\beta\omega^{\ast}}),
\end{displaymath}
\begin{displaymath}
\begin{split}
\mathcal {E}=-\frac{1}{V}\frac{\partial\ln Z}{\partial\beta}&=\Delta\varepsilon_{0}-T\frac{\partial\Delta\varepsilon_{0}}{\partial T}\\&-T^{2}\int
\frac{d^3p}{(2\pi)^3}\frac{\partial\ln(1-e^{-\beta\omega^{\ast}})}{\partial T}.
\end{split}
\end{displaymath}
Because the above formula is formally model independent, they automatically satisfy the fundamental thermodynamical relation between the pressure $\mathcal {P}(T)$ and the energy density
$\mathcal {E}(T)$
\begin{equation}
\mathcal {E}(T)=T\frac{d\mathcal {P}(T)}{d T}-\mathcal {P}(T).
\label{relation}
\end{equation}
For the same reason, we can generalize this formalism to the case of fields with spin. For the case of fermions, at zero chemical potential, the pressure and energy density are
\begin{displaymath}
\mathcal {P}=\frac{1}{\beta}\frac{\partial\ln Z}{\partial V}=-\Delta\varepsilon_{0}+4T\int\frac{d^3p}{(2\pi)^3}
\ln(1+e^{-\beta\omega^{\ast}}),
\end{displaymath}
\begin{displaymath}
\begin{split}
\mathcal {E}=-\frac{1}{V}\frac{\partial\ln Z}{\partial\beta}&=\Delta\varepsilon_{0}-T\frac{\partial\Delta\varepsilon_{0}}{\partial T}\\&+4T^{2}\int
\frac{d^3p}{(2\pi)^3}\frac{\partial\ln(1+e^{-\beta\omega^{\ast}})}{\partial T},
\end{split}
\end{displaymath}
which follow from the partition function
\begin{equation}
\ln Z=-V\beta\Delta\varepsilon_{0}+4V\int\frac{d^3p}{(2\pi)^3}\ln(1+e^{-\beta\omega^{\ast}}),
\end{equation}
where $\Delta\varepsilon_{0}$ is the effect of the classical background field in the quasi-fermions system. One can easily check that $\mathcal {P}(T)$ and $\mathcal {E}(T)$ automatically satisfy the 
thermodynamical consistency relation (\ref{relation}).

In Ref. \cite{25} Walecka deals with the infinity of the vacuum zero point energy in the mean field approximation at zero temperature. This is different from our case where the vacuum zero point energy is 
temperature dependent. Inspired by Walecka's approach, we have adopted the series expansion method to eliminate the temperature-dependent infinity of the vacuum energy at finite temperature. From this procedure
we have obtained a well-defined partition function and ensure that the thermodynamical consistency relation is automatically satisfied. 
 We also would like to point out that the above process of constructing quasi-particle model at finite temperature is general and can be directly generalized to the case of quasi-particle model at finite temperature and chemical potential. In this kind of system, the interaction between particles is related not only to the temperature but also to the chemical potential. In this case, we introduce a classical background field depending on the temperature and the chemical potential simultaneously to eliminate the temperature and chemical potential dependent infinity of the vacuum zero-point energy. In this way, we can obtain a well-defined partition function.

Here we would like to stress that a formal justification or any direct relation to QCD is a real challenge for simple quasi-particle models. The authors of Ref. \cite{32} point out in their paper that the missing transport peak in the spectral function invalidates quasi-particle models for strongly interacting systems. On the other hand, the flexibility of quasiparticle models allows one to describe very well the thermodynamical variables of the quark-gluon plasma, as shown by various authors in various examples \cite{9,8,10,20,14,11,15,12,22,13,31}. In this case, one needs to employ quasi-particle models to give predictions for more observables, as is done in Refs. \cite{31,Tian} (the authors of Ref. \cite{Tian} adopted the equation of state obtained in the quasi-particle model to calculate the mass-radius relation of the quark star and find that the result is consistent with most recent astronomical observation). Through comparison of theoretical results with experimental data, one can constantly improve the quasi-particle models. We think that the best means to pursue this goal is to focus on this particular model, making connections and comparisons with others whenever possible and helpful. 
According to this thought, we use our model to fit the latest lattice results of S. Bors\'{a}nyi $et$ $al$ for (2+1) flavor system \cite{30}. Considering the interacting plasma in thermodynamical equilibrium, we assume that it can be described as a system of massive quasi-particles, a picture arising asymptotically from the in-medium properties of the constituents of the plasma. Just as in Ref. \cite{20}, we write the effective mass of the quasi-particle as:
\begin{displaymath}
m_{i}^{2}(T)=m_{0 i}^{2}+\Pi_{i},
\end{displaymath}
where $m_{0 i}$ and $\Pi_{i}$ are the rest mass and thermal mass of the quasi-particle, respectively.
$\Pi_{i}$ are given by the asymptotic values of the gauge-independent hard-thermal-loop self-energies:
\begin{displaymath}
\Pi_{q}=2\omega_{q 0}(m_{0}+\omega_{q 0}),
\end{displaymath}
\begin{displaymath}
\omega_{q 0}^{2}=\frac{N_{c}^{2}-1}{16N_{c}}T^{2}G^{2}(T),
\end{displaymath}
\begin{displaymath}
\Pi_{g}=\frac{1}{6}\bigg(N_{c}+\frac{1}{2}N_{f}\bigg)T^{2}G^{2}(T),
\end{displaymath}
\begin{displaymath}
G^{2}(T)=\frac{48\pi^{2}}{(11N_{c}-2N_{f})\ln(\frac{T+T_{s}}{T_{c}/\lambda})^{2}},
\end{displaymath}
where $N_{c}$, $N_{f}$ stands for the color factor and the number of flavors, respectively, and $m_{0}$ is the rest mass of the quark. $T_{c}/\lambda$ is related to the QCD scale $\Lambda_{QCD}$. The quantity $G^{2}$ is to be considered as an effective coupling constant since it parametrizes all deviations of the exact spectral function from the ``strict-quasi-particle'' ansatz \cite{20}. Here we use $T_{c}=170 MeV$ as the transition temperature at vanishing chemical potential. Therefore, there is three free parameters: $\lambda$, $T_{s}$ and $B_{0}$ (which is the integration constant in the pressure) in our quasi-particle model. By fitting the results of lattice QCD for the pressure $\mathcal {P}$, the interaction measure $(\mathcal {E}-3\mathcal {P})$ and speed of sound squared $(C_{s}^{SB})^{2}$ we can determine the following values of these parameters: $\lambda=6.6$, $T_{s}$=$-0.88 T_{c}$ and $B_{0}^{1/4}=80 MeV$. As is shown in Figs. 1-3, our quasi-particle model can reproduce the lattice results well except the interaction measure $(\mathcal {E}-3\mathcal {P})$. According to the explanation of K.K. Szab\'{o} $et$ $al$ \cite{15}, the reason for this can be easily understood. First, we use the pressure $\mathcal {P}$ in our fitting procedure, therefore a better agreement is expected for this quantity than for the interaction measure $(\mathcal {E}-3\mathcal {P})$, which is a prediction of the model. Second, the interaction measure can be obtained as a partial derivative of the pressure:
\begin{displaymath}
\frac{\mathcal {E}-3\mathcal {P}}{T^{4}}=T\frac{\partial}{\partial T}\bigg(\frac{\mathcal {P}}{T^{4}}\bigg).
\end{displaymath}
A small difference in the pressure $\mathcal {P}$ can cause an evident deviation between the two results. In order to compare the results obtained in our model and those in earlier models, we also show the results calculated according to K.K. Szab\'{o} model proposed in Ref. \cite{15}. From these figures, it can be seen that our model can fit the recent lattice data better than earlier models. Moreover, in their model, in order to fit the lattice result, the gluon degeneracy factor is fixed to be $16.4_{-0.2}^{+0.3}$. So it is also a model parameter in earlier quasi-particle models. This is obviously unphysical. But in our model $d_{g}=16$, which is required by QCD. In other words, in our case there is no need to reserve this degree of freedom in the fitting procedure.

Here we would like to discuss the main difference between our model
and previous quasi-particle models. In the previous quasi-particle models, in order to satisfy the requirement of thermodynamical consistency, those authors in Refs. \cite{10,20,14,11,15,12,22,31} have to introduce an additional medium contributions which they also call $B(T)$. However, as discussed before, there exists a temperature-dependent infinity of thermal vacuum zero point energy and this makes the partition function
of the previous models ill defined. Whereas in our model, by means of
introducing a classical background field, we have successfully eliminated the temperature-dependent infinity of thermal vacuum zero point energy. This makes the partition function well defined and the thermodynamical consistency relation is automatically satisfied. 
 Here it is also interesting to compare these two $B(T)$ in the same graph for reference. The results are shown in Fig. 4. From Fig. 4 we can see that the overall trend of the curve of these two $B$ agrees qualitatively: as the temperature increases, they both decrease, although the rates are not the same.

\begin{figure}[t!]
\centerline{\includegraphics[width=0.9\columnwidth]{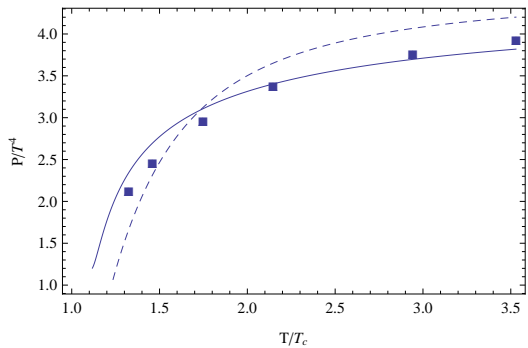}}
\caption{Lattice data of (2+1) flavor QCD for the pressure $P$ at $T>T_{c}$ normalized by $T^{4}$ \cite{30}(full boxes). The solid line corresponds to our quasi-particle model. The dashed line is the result obtained according to K.K. Szab\'{o}'s model in Ref. \cite{15}}
\label{P}
\end{figure}

\begin{figure}[t!]
\centerline{\includegraphics[width=0.9\columnwidth]{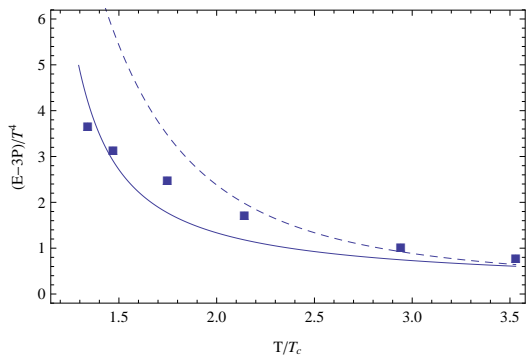}}
\caption{Lattice data of (2+1) flavor QCD for the interaction measure $\mathcal {E}-3\mathcal {P}$ at $T>T_{c}$ normalized by $T^{4}$ \cite{30}(full boxes). The solid line corresponds to our quasi-particle model. The dashed line is the result obtained according to K.K.Szab\'{o}'s model in Ref. \cite{15}.}
\label{e-3P}
\end{figure}

\begin{figure}[t!]
\centerline{\includegraphics[width=0.9\columnwidth]{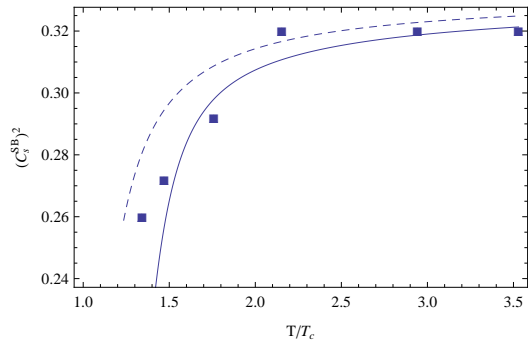}}
\caption{Lattice data of (2+1) flavor QCD for the speed of sound squared $(C_{s}^{SB})^{2}$ in the QCD plasma \cite{30}(full boxes). The solid line corresponds to our quasi-particle model. The dashed line is the result obtained according to K.K. Szab\'{o}'s model in Ref. \cite{15}.}
\label{Cs2}
\end{figure}

\begin{figure}[t!]
\centerline{\includegraphics[width=0.9\columnwidth]{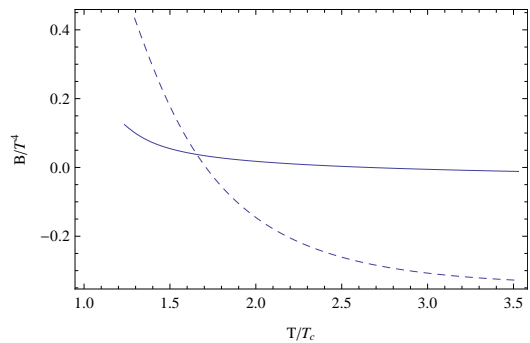}}
\caption{Additional medium contributions $B$ normalized by $T^{4}$ in K.K. Szab\'{o}'s model in Ref. \cite{15} (dashed line) and our classical background field $B$(solid line).}
\label{B}
\end{figure}

\section{Summary and conclusions}

To summarize, in the study of equilibrium statistical field theory, it is quite important to establish a well-defined partition function from which one can determine all the thermodynamical properties of the system. The primary goal of this paper is to build a well defined partition function for a quasi-particle system whose effective mass depends on the temperature. We first adopt the series expansion method inspired by J.D. Walecka's approach \cite{25} to deal with the $T$-dependent infinity of thermal vacuum energy by introducing a  classical background field into the effective Lagrangian. Through such a treatment we get a well-defined partition function. On the basis of this, following the standard ensemble theory we propose an improved quasi-particle model without any need of reformulation of statistical mechanics or thermodynamic consistency relation. Then, as an application of our quasi-particle model we apply it to the case of (2+1) flavor QCD at zero chemical potential and finite temperature. It is found that compared to the earlier model, our quasi-particle model show a nice agreement with the latest lattice simulation results of S. Bor\'{a}nyi $et$ $al$ \cite{30}. Moreover, in our model there is no need to reserve the gluon degeneracy factor as a model parameter in the fitting procedure, which is quite different with the previous models \cite{15}. Finally, We want to stress that our model framework is general and could be generalized to the case where the effective mass of the quasi-particle depends not only on the temperature but also on the chemical potential.

\section{acknowledgments}

This work is supported in part by the National Natural Science Foundation of China (under GrantNos 10775069, 10935001 and 11075075), the Research Fund for the Doctoral Program of Higher Education
(under Grant No 200802840009) and a project funded by the Priority Academic Program Development
of Jiangsu Higher Education Institution.

\end{document}